\documentclass[prd,tightenlines,nofootinbib,amsfonts,amssymb,amsmath,color,11pt]{revtex4}
\usepackage{graphicx}
\usepackage{color}

\begin{document}
\newcommand{\etal}{{\it et al.}}
\newcommand{\bx}{{\bf x}}
\newcommand{\bn}{{\bf n}}
\newcommand{\bk}{{\bf k}}
\newcommand{\dd}{{\rm d}}
\newcommand{\dslash}{D\!\!\!\!/}
\def\ga{\mathrel{\raise.3ex\hbox{$>$\kern-.75em\lower1ex\hbox{$\sim$}}}}
\def\la{\mathrel{\raise.3ex\hbox{$<$\kern-.75em\lower1ex\hbox{$\sim$}}}}
\def\beq{\begin{equation}}
\def\eeq{\end{equation}}
\def\be{\begin{equation}} 
\def\ee{\end{equation}}
\def\bea{\begin{eqnarray}}
\def\eea{\end{eqnarray}}

\def\C{{\cal C}}
\def\cleeu{{\tilde \C}_l^{EE}}
\def\clteu{{\tilde \C}_l^{TE}}
\def\clttu{{\tilde \C}_l^{TT}}
\def\a{\alpha}
\def\im{\eta^{-1}}

\def\tr{{\rm tr}}


\vskip-2cm
\title{A small  cosmological constant  from  Abelian symmetry breaking}

\author{  Gianmassimo Tasinato}
\affiliation{
 Institute of Cosmology \& Gravitation, University of Portsmouth, Dennis Sciama Building, Portsmouth, PO1 3FX, United Kingdom\\
}
\vspace*{2cm}
 
\begin{abstract}

We investigate  some cosmological consequences of a vector-tensor theory   
  where an Abelian symmetry 
  in the vector sector  is slightly broken by a mass
  term and by ghost-free derivative self-interactions.
  When studying cosmological expansion
   in the presence of large bare cosmological constant $\Lambda_{cc}$, we find that the theory admits  branches of
   de Sitter
    solutions in which
   the scale of the Hubble parameter is inversely proportional to a power of  $\Lambda_{cc}$. Hence,
    a  large value of $\Lambda_{cc}$ leads to a small size  for the Hubble scale.
         In an appropriate
  limit, in which the symmetry breaking parameters are small, the theory recovers the Abelian symmetry plus an 
  additional Galileon symmetry acting on the longitudinal vector polarization.       The approximate 
  Galileon symmetry  
   can make    the structure
    of this theory 
     stable
     at the energy scales 
      we are interested in. We also analyze  the  dynamics of 
     linearized 
    cosmological
 fluctuations around the de Sitter  
    solutions, showing that 
     no manifest instabilities arise, and that the transverse vector polarizations become massless around these configurations.

\vspace{1cm}

\end{abstract}

 \date{April 2014}
 \maketitle
 
 \section{Introduction}

  One of the main 
  motivations for considering modifications of Einstein's general relativity  (GR)
is to understand 
 why the universe is accelerating today.   Cosmological acceleration might be due to the dynamics
   of gravity itself, that at very large distances deviates from the predictions of GR
    in such a way to provide accelerated expansion with no need of energy momentum tensor. This phenomenon is dubbed self-acceleration.
         The   
      DGP model \cite{Dvali:2000hr}
  is  among the most studied examples, but various other proposals have been explored in the literature.
  Sometimes, the self-accelerating branches of cosmological solutions in these set-ups are plagued by instabilities.   
   Nevertheless, the exploration of modified gravity models is certainly worthwhile for better understanding  
  subtleties associated with  the dynamics of 
   gravity,   and for  
  suggesting new theoretical  ideas to address 
  the   dark energy problem. 
 %
  See \cite{Clifton:2011jh} for a comprehensive review on this subject.
    
    \smallskip
    
 Yet, most of these theories do not attempt to solve the classic  cosmological
 constant problem, since they  do not try to explain why 
  the bare  vacuum energy does not curve the space-time.
   In absence of symmetries, one would expect that quantum effects give contributions to vacuum energy
     that scale as the fourth power of the 
   cut-off of the theory. In the theory of general relativity, without
    imposing additional symmetries beyond diffeomorphism invariance, the natural cut-off is the Planck mass. Hence, 
      the expected size of the cosmological constant would scale as Planck mass to the fourth, that is
    120 orders of magnitude larger than  the measured dark energy scale today. 
      See \cite{Weinberg:1988cp} for a classic review on the cosmological constant problem, and \cite{Polchinski:2006gy,Burgess:2013ara} for
   recent perspectives on this topic. The   work \cite{Weinberg:1988cp} also contains a powerful result, the so called Weinberg no-go theorem, 
   that seems to forbid adjustment mechanisms for reducing the size of the cosmological constant; see however \cite{Aghababaie:2003wz,Charmousis:2011bf,Kaloper:2013zca,Batra:2008cc,deRham:2010tw,deRham:2011by} for some
    interesting recent
   proposals for avoiding  Weinberg's arguments. 
   
   The cosmological constant problem might   be due to the particular structure  and symmetries of GR.
    Enriching the theory with new symmetries,
 as for example supersymmetry or   conformal symmetry, might  help to reduce the size of the problem by many orders
 of magnitude. On the other hand, at least within the  energy scales probed by current laboratory experiments,  
   these  additional  symmetries are  not manifest, hence they cannot be effective for reducing 
  to acceptable levels the size
 of the cosmological constant.

   \smallskip
In this work, we point out that 
   a vector field with an approximate 
Abelian 
symmetry might shed some light on why the cosmological constant does not curve the space.
 We elaborate on the theory presented in \cite{Tasinato:2014eka}, 
 in which the Abelian symmetry of a vector coupled to gravity is broken by a mass
  term and derivative self-interactions. (See also \cite{Kim:2013jka} 
   for  a proposal to describe Dark Energy by means of a theory breaking an Abelian symmetry.)  
  The theory is ghost-free and propagates three degrees of freedom
  in the vector sector: the two transverse vector components and a longitudinal scalar field. In an appropriate
  limit, in which the symmetry breaking parameters are small, the theory recovers an Abelian symmetry plus a  
  Galileon symmetry acting on the longitudinal scalar mode. We study the behavior of cosmological
  expansion in this set-up in the presence of a large bare cosmological constant $\Lambda_{cc}$.
  We show that, interestingly, there are branches of de Sitter cosmological solutions in which the square of the Hubble
  parameter is inversely proportional to a power of $\Lambda_{cc}$. Hence, the larger  the bare cosmological constant
   is, 
  the smaller the value of the Hubble constant, and parameters can be chosen in such a way that the resulting Hubble scale
  is of the order of the observed  dark energy scale. 
  The existence of these cosmological solutions depends on the particular structure of our vector theory: exploiting  approximate 
  Galileon symmetries in the longitudinal sector, we argue that
 this structure is stable at the energy  scales we are interested in. We also analyze  the  dynamics of cosmological
 fluctuations around these de Sitter  
    solutions, showing that 
     no manifest instabilities arise.
        
   
     For simplicity, in this work we will not discuss the role of  standard matter in cosmological expansion, and focus
     on the dynamics of the vector. Also, 
         we will not be specific on the   nature of the vector we consider.  It could be the observed photon;
     or, alternatively, it could be another light vector field not belonging to the Standard Model.
    Contrarily to other hypothetical symmetries,  an Abelian gauge symmetry acting on the photon field seems  to be a fundamental symmetry realized in nature, at least to a level that did not manifest  any appreciable violation  so far: see \cite{Goldhaber:2008xy}
     for an excellent  review on photon mass limits. Yet, the study
     the cosmological consequences of tiny violation of Abelian gauge symmetries 
     is interesting, since being associated with  a
      long-range force    might have an important role for shaping  
     the structure of  our universe at very large scales.

 Various vector tensor theories have been developed so far with  interesting 
cosmological applications. The first ones are  \cite{Will:1972zz,Hellings:1973zz}, while more recent well studied scenarios are
the Einstein- Aether theory \cite{Jacobson:2000xp} and the TeVeS covariantized version of MOND \cite{Bekenstein:2004ne}.
 For more recent  examples see e.g. \cite{Jimenez:2013qsa},
and more generally see \cite{Clifton:2011jh} for a comprehensive review on the subject with a complete list of references to the relevant literature. 
The specific 
feature of 
 our approach is the emphasis on symmetry arguments for discussing our theory, in particular the explicit connections with Galileons. This fact can allow us to keep the structure of our set-up under control 
 within 
  the cosmological scales we are interested in.

 The work is organized as follows. In Section \ref{sec-first} we present 
 the set-up under consideration, determining homogeneous cosmological solutions. In Section {\ref{sec-perts}} we study the dynamics
 of cosmological perturbations around these solutions. In Section {\ref{sec-DE}} we apply our findings to analyze the size of the dark energy scale, showing that there are interesting branches of solutions in which the Hubble scale is inversely proportional to the bare cosmological
 constant. We conclude in Section \ref{sec-concl}.

 \section{The system under consideration}\label{sec-first}
 We consider a theory of vector fields 
 coupled to gravity, described by an action
 \be
 {\cal S}\,=\,\int d^4x\,\sqrt{-g}\,\left[\frac{M_{*}^2}{2}\,R-\frac14 F_{\mu\nu} F^{\mu\nu}-{\cal L}^{cov}_{(0)}-{\cal L}^{cov}_{(1)}
 -{\cal L}^{cov}_{(2)}-\,\Lambda_{cc} \right]
 \,,
 \label{inact}
 \ee 
 where $  {M_{*}^2}\,R/{2}$
  is the  Einstein-Hilbert term weighted by the square of Planck scale, $- F_{\mu\nu} F^{\mu\nu}/4$ is the 
  standard 
  kinetic term for a vector potential $A_\mu$,
   $\Lambda_{cc}$ is a bare cosmological constant, and the
   vector 
   interactions ${\cal L}^{cov}_{(i)}$ that break the 
   Abelian
    symmetry  
  are defined as
  \bea\label{l0co}
{\cal L}^{cov}_{(0)}&=&m^2\,A_\mu A^\mu\,,
\\\label{l1co}
{\cal L}^{cov}_{(1)}&=&{\beta_1}\,A_\mu A^\mu\,
\left( \nabla_\rho A^{\rho}\right)
  \,,
\\\label{l2co}
{\cal L}_{(2)}^{cov}&=&\frac{\beta_2}{m^2}\,A_\mu A^\mu\,\left[
\left( \nabla_\rho A^{\rho}\right) \left( \nabla_\nu A^{\nu}\right) 
-\left( \nabla_\rho A^{\nu}\right) \left( \nabla^\rho A_{\nu}\right) 
 -\frac14\,R\,A_\sigma A^\sigma\right]\,.\label{cova2}
\eea 
This is 
  the minimal ghost-free Lagrangian studied in \cite{Tasinato:2014eka}  that 
  couples 
  a vector field with gravity, and leads 
     to cosmological solutions with  interesting features
 that  we are going 
  to analyze. ${\cal L}_{(0)}^{cov}$ is a Proca mass term, while ${\cal L}_{(1,\,2)}^{cov}$ are 
   ghost-free 
  derivative self-interactions. 
    The structure of the self-interactions is chosen in such a way that the 
   equation  of motion for the 
   time-component $A_0$ of the vector field does  not contain time derivatives. Hence the equation for 
   $A_0$ is a constraint  equation that fixes one degree of freedom, and the theory propagates
   only three degrees of freedom in the vector sector \footnote{A similar set of interactions is studied in \cite{Gripaios:2004ms}, but 
   the coupling of the Ricci scalar
     with gravity is not taken in due account, rendering the theory  not ghost-free when dynamical gravity is considered.}.
    The Lagrangian   can be further generalized maintaining the   ghost free condition, see for example \cite{Heisenberg:2014rta}:
    however the minimal form of the action that we consider is particularly interesting for us because of 
 symmetry    properties   that we will exploit in what follows.   
 We add
  a bare cosmological constant $\Lambda_{cc}$ in order to analyze how cosmological solutions
  depend on its size, but 
  for simplicity    we will not explicitly  discuss couplings with standard matter -- although
   we will comment on this topic from time to time.

\smallskip

   In general, this theory does not have any symmetry besides the diffeomorphism invariance of GR:
    indeed, the 
     Lagrangians  ${\cal L}^{cov}_{(i)}$ 
       break the $U(1)$ Abelian gauge symmetry
    $A_\mu\,\to\,A_\mu+\partial_\mu \xi$. 
    On the other
   hand, 
 as explained  in \cite{Tasinato:2014eka}, there exists
 a  limit in which, by neglecting
  gravity and taking   
  small   values for $m$, $\beta_i$ 
 the theory  acquires  
 Abelian and Galileon symmetries  acting on the transverse and longitudinal vector polarizations.
  This limit is made particularly manifest by adopting a St\"uckelberg approach, and supplementing the Lagrangians ${\cal L}_{(i)}^{cov}$
  of eqs (\ref{l0co}-\ref{l2co}) with  interactions to a new scalar $\pi$, introduced   in such a way to  obtain a
  gauge-symmetric
    theory 
  
  \bea
{\cal L}^{cov}_{(0)}&=&m^2\,\left(A_\mu+\frac{1}{\sqrt{2}\, m}\partial_\mu \pi\right)\,\left( A^\mu+\frac{1}{\sqrt{2}\, m}\partial^\mu \pi\right)\,
\\
{\cal L}^{cov}_{(1)}&=&{\beta_1}\,\left(A_\mu+\frac{1}{\sqrt{2}\, m}\partial_\mu \pi\right)\,\left( A^\mu+\frac{1}{\sqrt{2}\, m}\partial^\mu \pi\right)\,
\left( \nabla_\rho A^{\rho}+ \frac{1}{\sqrt{2}\, m}\,\Box\, \pi\right)
  \,,
\\
{\cal L}_{(2)}^{cov}&=&\frac{\beta_2}{m^2}\,\left(A_\mu+\frac{1}{\sqrt{2}\, m}\partial_\mu \pi\right)\,\left( A^\mu+\frac{1}{\sqrt{2}\, m}\partial^\mu \pi\right)\,\,\nonumber\\
&\times&\Big[
\left( \nabla_\rho A^{\rho}+ \frac{\Box\, \pi}{\sqrt{2}\, m}\right)
\left( \nabla_\nu A^{\nu}+ \frac{\Box\, \pi}{\sqrt{2}\, m}\right)
-
\left( \nabla_\rho A^{\nu}+ \frac{\nabla_\rho \partial^\nu\, \pi}{\sqrt{2}\, m}\right)
\left( \nabla_\nu A^{\rho}+ \frac{\nabla_\nu \partial^\rho\, \pi}{\sqrt{2}\, m}\right)
%
 \nonumber\\&&\hskip0.5cm-\frac14\,R\,\left(A_\sigma+\frac{1}{\sqrt{2}\, m}\partial_\sigma \pi\right)\,\left( A^\sigma+\frac{1}{\sqrt{2}\, m}\partial^\sigma \pi\right)\Big]\,\label{cova2a}
 \,.
\eea 
The field $\pi$ plays the same physical role of the longitudinal vector mode in the original theory; on the other hand,
now  the interactions listed  above preserve the Abelian gauge symmetry $A_\mu\to A_\mu+\partial_\mu \xi$
  at the price of introducing the degree of freedom $\pi$ that simultaneuosly transforms as $\pi\to\pi-\sqrt{2} m\,\xi$. 

\smallskip

  Let us now consider the following   limit that decouples the dynamics of transverse and longitudinal vector modes:
 \bea
 m\to 0\hskip0.5cm,\hskip0.5cm 
   \beta_1&\to&0\hskip0.5cm,\hskip0.5cm  \beta_2\,\to\,0
  \,.\hskip0.5cm\hskip0.5cm\label{declimit}
\eea
 This limit 
must be taken in a special way: indeed
 we require  the following conditions for
  the ratios among the previous parameters in the limit in which  they go to zero
\bea
 \frac{\beta_1}{m^3}&\equiv&\frac{\hat{\beta}_1}{\Delta^3}\,=\,{\rm fixed}\hskip0.5cm,\hskip0.5cm\frac{\beta_2}{m^6}\,\equiv\,\frac{\hat{\beta}_2}{\Delta^6}\,=\,{\rm fixed}\,.
 \label{defratio}\eea
 In the previous expression,
  $\hat{\beta}_{1,\,2}$ are dimensionless constants, while $\Delta$ is a quantity of dimension of a mass. In this decoupling limit, when neglecting  gravity,
 the theory enjoys an Abelian symmetry acting on the vector sector only $A_\mu\,\to\,A_\mu+\partial_\mu \xi$, and (as we will
 see in a moment) a Galileon symmetry
 $\pi\to\pi+c+b_\mu x^\mu$
  for the scalar sector $\pi$, with a strong
 coupling Galileon scale  set by $\Delta$. \footnote{At least if the $\hat{\beta}_i$ are of order one: if these parameters are small, the strong coupling scale can be higher.}  The fact that we recover symmetries in the limit in 
 which
  $m$, $\beta_1$ 
  and  $\beta_2$  are going to zero implies that choosing these parameters small is technically natural in the 't Hooft sense \cite{'tHooft:1979bh}: at the quantum level, we do not expect large contributions  
  to their size 
  proportional to powers of the cut-off, but only corrections
  proportional to the size of the small parameters themselves.

 

 \smallskip
 
 Let us return to discuss the sector of the theory associated with the longitudinal vector polarization. 
   In the decoupling limit (\ref{declimit}), (\ref{defratio}), the effective Lagrangian density for the vector longitudinal
 polarization $\pi$ reads
 
 \be
 {\cal L}_{\pi}^{dec}\,=\,-\frac12\,\left(\partial \pi \right)^2+\frac{\hat{\beta}_1}{\Delta^3}\,{\cal L}_{(3)}^{gal}
 +\frac{\hat{\beta}_2}{\Delta^6}\,{\cal L}_{(4)}^{gal}
 \ee
 where ${\cal L}_{(3,\,4)}^{gal}$ are the cubic and quartic Galileon Lagrangians discussed in \cite{Nicolis:2008in};
  see
 \cite{Tasinato:2014eka} for more details.   
   In this limit we can exploit the non-renormalization theorems of Galileon interactions \cite{Luty:2003vm,Nicolis:2004qq,Nicolis:2008in}. Our choice
 of operators in (\ref{inact}), although special, is technically natural since quantum corrections do not spoil the structure of  
 the  theory when the {\it vev} of   $\pi$  is    of  order $\Delta$.
 On the other hand, 
the coupling with  gravity breaks the Galileon symmetry (while it preserves the Abelian vector symmetry, and by construction maintains
equations of motion of second order).  In a cosmological setting,  corrections associated with gravitational effects are expected to be small in the limit in which
the strong coupling scale $\Delta$ is larger than the Hubble scale of the  space-time under consideration,  $H\ll\Delta$
  \cite{Burrage:2010cu,deRham:2012az}. At the same time, 
  Planck suppressed operators induced by quantum corrections are expected to be 
      negligible   
  in the limit in which the strong coupling  scale is  much smaller than the gravitational cut-off, $\Delta\ll M_\star$.
  Further corrections to the structure of the Lagrangians contained in eq. (\ref{inact}) can arise since
    finite values for the vector mass and self-interaction coefficients $m$, $\beta_i$ break the Galileon symmetry by coupling longitudinal and transverse degree of freedom -- on the other hand as argued above  these couplings can be sufficiently suppressed by  choosing (technically natural) small values for these parameters.
  Finally,   additional  corrections   can arise when coupling the theory to standard matter -- unless this is done 
   in a careful  way such to maintain   the above symmetries to a certain extent.  
  This is an important subject, since couplings with matter will also govern the scale of the  typical {\it vev} for the vector longitudinal polarization $\pi$, offering the possibility to develop screening effects analogous to the gravitational Vainshtein mechanism \cite{Vainshtein:1972sx} (see also 
   \cite{Tasinato:2013oja} for a  realization of a Vainshtein mechanism in a vector-scalar-tensor theory of gravity).
   This subject
  has been partially explored in \cite{Tasinato:2014eka}, but will be further developed 
  elsewhere.

   \smallskip
   
These considerations suggest that,  when limiting
 our interest to cosmological space-times with  Hubble parameter smaller 
  than the strong coupling scale $\Delta$ for the vector longitudinal interactions,
 %
 it is possible to maintain   a sufficient degree of symmetry to protect in a technically natural way the
structure of our theory.  This is a feature that will have important implications for our arguments. Let
us then pass to analyze, for the remaining of this section, the   
    homogenous cosmological evolution associated with the above theory, extending results 
    first presented in  
   \cite{Tasinato:2014eka}.  
 The Ans\"atze for the FRW background metric and vector profiles that we consider are 
\bea
d s^2&=&-d t^2+a^2(t)\,\delta_{ij}\,d x^i d x^j \,,
\\
A_\mu&=&(A_0(t),\,0,\,0,\,0)\,.
\eea
 Notice that the homogeneous vector profile does not break spatial isotropy. 
 The homogenous  equation of motion for the vector component 
is a constraint equation, that reads
\be
A_0\,\left(m^4-3\,\beta_1 m^2\,A_0\,H+9 \,\beta_2\,A_0^2\,H^2 \right)
\,=\,0\,,
\ee
where $H\,=\,\dot{a}/a$ is the Hubble parameter. The above algebraic equation, 
a part from the  solution $A_0=0$ (that does not lead to interesting
 cosmological expansion),    admits the solutions 
\bea
A_0&=&\frac{c_\pm\,m^2}{H}\label{solfA0}
\,,\\
c_\pm&=&\frac{\beta_1\pm\sqrt{\beta_1^2-4\beta_2}}{6\,\beta_2}
\,,\eea
where from now on we set $m^2>0$, $\beta_1>0$. Hence we learn that, in a homogeneous FRW setting, the 
constraint equation for $A_0$ leads to a profile for this field that is inversely proportional to the Hubble parameter. 
In order to ensure real values for $c_\pm$, we will impose $\beta_2\le \beta_1^2/4$.  
For  handling more easily our formulae, it is convenient to make the following parameter redefinition:
\bea
\beta_2&=&\frac{(1-\gamma^2)\,\beta_1^2}{4}\label{b2def}
\,,\\\label{fdlcc}
 \Lambda_{cc}&=&\frac{m^3\,M_{*}}{3\,\beta_1}\,\lambda\label{defcc}
\,,\eea
where $\gamma$, $\lambda$ are dimensionless quantities. This implies that we trade $\beta_2$
for $\gamma$; in terms of the parameters $\hat{\beta}_i$ of eq. (\ref{defratio})  (useful to investigate the decoupling limit 
 (\ref{declimit})
 in which we recover
Abelian and Galileon symmetries) we can write
\bea
\hat{\beta}_2&=&\frac{(1-\gamma^2)\,\hat{\beta}_1^2}{4}\,.
\eea
That is, $\gamma$ is not affected by the limit (\ref{declimit}). 
Notice that eq (\ref{defcc}) means that we are going to  
   use the dimensionless parameter $\lambda$ to quantify the size of the cosmological
constant. 
 For the moment  we do not impose any requirement on the size of  $\lambda$, that could also be very large (so to drive $\Lambda_{cc}$
  up to the gravitational  cut-off
 of our theory). How  the cosmological expansions depends on
  $\lambda$
 and then  on   the bare cosmological constant $\Lambda_{cc}$ will be the main topic of Section \ref{sec-DE}.  
   Using eq. (\ref{b2def}), the parameter $c_\pm$ reads 
\be
c_{\pm}\,=\,\frac{2}{3\,\beta_1\,\left( 1\mp|\gamma|\right)}\,.
\ee
Notice  that the $c_{\pm}$ are distinguished by the sign in front of  $|\gamma|$. Hence  in what follows, without
lack of generality, we will write in
an unified way these two branches 
 as 
\be
c_0\,=\,\frac{2}{3\,\beta_1\,\left( 1+\gamma\right)}\,,
\ee
 and allow for an arbitrary  sign of $\gamma$. 
  In terms of these parameters, the Einstein equations 
  reduce to the following condition for the Hubble parameter
  \bea
  0&=&H^2\left( -2\, \Lambda_{cc} 
  +6 \,H^2 \,M_\star^2-2m^2\,A_0^2
+12\,\beta_1\,H\,A_0^3-\frac{45\,\beta_2}{m^2}\,H^2\,A_0^4  \right)
  \eea
  and substituting  (\ref{solfA0}) in the previous equation we find
     two allowed 
branches of non-vanishing constant solutions for the Hubble parameter
\bea\label{H2brs}
H^2_\mp&=&\frac{m^3}{18\,\beta_1\,M_{*} }\,\left[
{\lambda\mp\sqrt{\lambda^2-\frac{24 (1+3\gamma)}{(1+\gamma)^3}
}}
\right]
\,.\eea
So, we learn that the higher order  self-couplings of the vector, controlled by the interaction
Lagrangians (\ref{l0co}-\ref{l2co}),  
 switch on a non-trivial time-dependent profile for the component $A_0(t)$, that  
  drives  cosmological expansion.
Choosing parameters such that  the right hand side of eq. (\ref{H2brs}) is positive, the resulting cosmological
evolution corresponds to a de Sitter universe with constant Hubble rate. 
  To have a positive argument for the square root in  (\ref{H2brs}), we impose the following condition
 on the dimensionless parameters $\lambda$ and $\gamma$:
 \be\label{coslam}
 \lambda^2\ge\frac{24 (1+3\gamma)}{(1+\gamma)^3}\,.
 \ee 
 After imposing (\ref{coslam}), 
   we can distinguish two options to obtain a positive value for the square of the Hubble parameter:
 \begin{enumerate}
 \item
  If $\lambda$ is positive, the positive branch $H_+$ is always well defined, in the sense that $H_+^2$ is
  positive. 
In the case of negative branch $H_-$, moreover, to have a positive $H_-^2$ we have to additionally demand 
\be
\frac{ 1+3\gamma}{(1+\gamma)^3}
\,\ge\,0\,\hskip0.5cm \to \hskip0.5cm
  \gamma
\le-1\,\,{\rm or}\,\,\gamma\ge-\frac13
\,.
\ee
\item If 
$\lambda$ is negative, the negative branch $H_-$ is never well defined. 
In the case of positive branch $H_+$, moreover, to have a positive $H_+^2$ we have to additionally demand 
\be
\frac{ 1+3\gamma}{(1+\gamma)^3}
\,\le\,0\,\hskip0.5cm \to \hskip0.5cm
 -1\le\gamma
\le-\frac13
\,.
\ee
\end{enumerate}
The negative branch $H_-$ for the Hubble parameter in eq  (\ref{H2brs}) appears  particularly interesting, since the minus sign inside the square parenthesis in (\ref{H2brs}) 
 compensates possibly large contributions associated with the parameter $\lambda$ (appearing in the expression (\ref{fdlcc}) for
 the bare cosmological constant).
   We  will explore in detail    this feature   
  in section \ref{sec-DE}.  In the next section, instead, we analyze the behavior of cosmological fluctuations around the homogeneous
  configurations we have determined.

\section{Dynamics of linearized perturbations}\label{sec-perts}

The dynamics of cosmological fluctuations around the previous background solutions can be analyzed straightforwardly. 
 It is convenient to split perturbations into tensor, vector, and scalar components 
 with respect to the spatial sections of the FRW background geometry,  and implement an  ADM
 approach. 
 Let us start  counting  the available degrees of freedom (dofs). We start with 
ten dofs in our metric, plus four dofs for the vector. The theory  respects diffeomorphism  invariance of GR, that removes four dofs. Moreover, we still have the four constraints of GR, plus one constraint of the vector action (associated with the equation of motion for  $A_0$ \cite{Tasinato:2014eka}):  these constraints remove five dofs. In total, we remain with five dofs: two transverse traceless tensor components, two transverse vector
components, one scalar component.

\smallskip

  Using diffeomorphism invariance, a gauge can be selected for metric fluctuations such that we can write
\be\label{mefl}
 g_{\mu\nu}\,=\,
 -\left(1+ N\right)^2 d t^2+ a^2(t)\,
 \, \left(\,e^{2\,\zeta}\,\delta_{ij}+h_{ij}\right)\left(d x^i+N^i d t\right)\left(d x^j+N^j d t\right)
\,,
 \ee
where $N$, $N_i$, $\zeta$, $h_{ij}$  are small fluctuations around the background
FRW solution characterized by a 
  scale factor $a(t)$. $N$ and $N_i$ are the lapse and shift perturbations, whose
 linearized  equations of motion provide the  
the Hamiltonian and momentum constraints  
 of GR. $\zeta$ is a scalar fluctuation, while $h_{ij}$
is a tensor fluctuation satisfying a
transverse-traceless  condition $h_{ii}=0$, $\partial_i h_{ij} =0 $. Vector fluctuations in the metric are set to zero by  gauge choice.  
 These conditions  do not completely fix the gauge:
  at zero momentum we have the additional freedom to shift the 
   time coordinate $t\,\to\,t+{\epsilon}(t)$ and rescale the spatial coordinates $x^i\,\to\,(1+q)\, x^i$ by small quantities  \cite{Weinberg:2003sw}.

For what respect the  vector sector with slightly broken Abelian symmetry,
 it is convenient to use the St\"uckelberg approach, and write $A_\mu$ as

   \be\label{vefl}
 A_\mu\,=\,({A}_0+\dot{\pi},\,\hat{A}_i^T+\partial_i \pi)\,.
 \ee
 
 \noindent
 ${A}_0$ corresponds to the homogeneous background solution discussed in the previous section, while $ \hat{A}_i^T$ and $\pi$ 
  play the role of 
  transverse vector and 
 scalar fluctuations. 
     We introduced a Stuckelberg field $\pi$ 
    restoring the Abelian invariance, that we use  to choose an Abelian gauge where the fluctuation $\hat{A}_0=0$. 
  Notice that in this description we apparently have six dofs, instead of the expected five, so 
  one of the scalars is actually  a gauge mode.
   Since  in our system gravity is non-minimally coupled with the vector Lagrangian,   we expect that the  dynamical scalar dof    will be a mixture between the scalars $\zeta$ and $\pi$. 
    It is convenient to work with both these scalars to start with,
      and leave the constraint equations to determine
      the   scalar combination playing a physical role in what follows.

\subsection{Tensor perturbations}
The quadratic action for tensor perturbations can be found straightforwardly. Notice that, when $\beta_2$ is non-vanishing (i.e. $|\gamma|\neq1$, see eq (\ref{b2def})) the vector is non-minimally coupled with the metric: hence the gauge field background value $A_0$
 `renormalizes' the Planck mass.
  Indeed, one finds the following effective Lagrangian density at quadratic level for the tensor 
  fluctuations $h_{ij}$
  \be
  {\cal L}^{(quad)}_{tens}\,=\,\frac{M_{\pm}^2}{2}\,{\cal L}^{(quad)}_{EH}
\,,  \ee
  where ${\cal L}^{(quad)}_{EH}$ is the expansion of the Einstein-Hilbert action at quadratic order, 
  and the effective Planck scale is given by
\be\label{fmeff}
M_{\pm}^2\,=\,M_{*}^2\,\left( 1-\frac{3\,\beta_2\,c_0^4\,m^6}{2\,H_{\pm}^4\,M_{*}^2}\right)\,.
\ee
In the previous formula, the $\pm$ denotes the choice of branch of background solutions for the Hubble parameter
in eq. (\ref{H2brs}).  Using
the results of Section \ref{sec-first},  
$M_{\pm}$ can
 be expressed as 
\bea\label{pmeff}
M^2_{\pm}&=&\left( 1+\frac{24\,\left( 1+\gamma\right)}{
\left( \gamma-1\right)^3
\,\left( \lambda\pm\sqrt{\lambda^2-\frac{24(3\gamma-1)}{(\gamma-1)^3}
}
\right)^2
}\right)\,M_{*}^2\,.
\eea
In order to have a consistent set-up, we impose  $M^2_{\pm}>0$.  
Hence, if $\gamma\neq-1$ the effective Planck scale depends on the value of the cosmological constant, since it explicitly depends
on $\lambda$, the parameter that controls $\Lambda_{cc}$ (see eq (\ref{defcc})).
  Let us point out  that the quantity $M^2_{\pm}$  of eq (\ref{pmeff}) can be interpreted as 
 parameterizing   the   self-coupling scale of gravitational interactions.  On the other hand,  if vector fields directly couple
  with standard matter, they can also have a role  in determining   
   the effective coupling of gravity with any additional  matter content.

\subsection{Vector perturbations}

Also vector fluctuations are not difficult to deal with. By splitting the metric shift vector $N_i\,=\,N_i^T+\partial_i \psi$, with $N_i^T$
 the transverse components and $\psi$ the longitudinal part, 
  the momentum constraint imposes $N_i^T=0$.  
Interestingly, a straightforward calculation shows that the mass
of the transverse vector fluctuations $\hat{A}_i^T$ {\it exactly vanishes} around the  background cosmological configurations
we are considering: at quadratic order, the Lagrangian density for the vector fluctuations $\hat{A}_i^T$ only contains  the standard
kinetic terms:
\be
 {\cal L}^{(quad)}_{vec}\,=\,-\frac{1}{4}
 \,F_{\mu\nu} F^{\mu\nu}
\,. \ee
 Hence, we are dealing with a transverse vector fluctuations with healthy kinetic terms and zero mass (although  the
 transverse polarizations
 acquire 
    interactions with the longitudinal component at third order in perturbations). 
   If we interpret   the vector we are dealing with  as    the  photon, 
  this implies that the usual constraints on the
  photon mass 
  do not directly apply in the present context,
    since the vector is massless. It would  be interesting to study in detail the 
 phenomenological consequences of 
the  higher order interactions associated with Lagrangians ${\cal L}_{(1)}^{cov}$
and  ${\cal L}_{(2)}^{cov}$, that can lead to screening mechanisms analogous to the gravitational
Vainshtein  mechanism.  This will be the subject
of a future work.

\subsection{Scalar perturbations}
The analysis of scalar vector fluctuations is
also straightforward. 
 The Hamiltonian  and momentum constraint equations, using also the
gauge freedom left at zero momentum, 
  provide the following
conditions (recall that $\psi$ is the longitudinal scalar part of the shift perturbations $N_i\,=\,N_i^T+\partial_i \psi$)
\bea
N&=&\frac{H_{\pm}}{c_0\,m^2}\,\dot{\pi}\,,\\
\psi&=&-\frac{H_{\pm}}{c_0\,m^2}\,\pi\,,\\
\zeta&=&\frac{H^2_{\pm}}{c_0\,m^2}\,\pi\,. \label{zetaeq}
\eea
So all the quantities can be expressed in terms of the St\"uckelberg scalar $\pi$, physically corresponding to
the  vector longitudinal polarization,  
and as expected one ends with a single (potentially) dynamical scalar fluctuation.
 However, by imposing the above constraints to the 
  action of quadratic scalar fluctuations around the homogeneous background, one ends with a total derivative signaling  a trivial
  dynamics for the scalar $\pi$:
  \be
 {\cal L}^{(quad)}_{scal}\,=\,0\,.
 \ee
 
   This implies that at quadratic level our action propagates only four
  dofs around the homogeneous de Sitter background, in the tensor and vector sectors.  It remains to be checked 
   whether higher order contributions in the fluctuations induce 
   instabilities in the scalar sector, analogously to what happens, for example, in the vector sector of massive gravity \cite{Koyama:2011wx,Tasinato:2012ze}.
    On the other hand, let us point out that this problem, if it really exists,  might be cured by suitable couplings with matter sector, that might be able 
    to induce healthy kinetic terms for $\pi$. We postpone a detailed analysis of this issue for future work, and we pass
    to discuss the consequences of our findings so far for the dark energy scale.  

\section{The dark energy scale}\label{sec-DE}

We have seen that coupling our vector Lagrangian with gravity we are able to find cosmological solutions corresponding 
to a de Sitter universe. We include  the effect  of a bare cosmological constant $\Lambda_{cc}$, that in 
 eq (\ref{defcc})  
 we parameterized in terms of the available
parameters as
\bea
 \Lambda_{cc}&=&\frac{m^3\,M_{*}}{3\,\beta_1}\,\lambda\label{defcc2}\,.
 \eea
 In the previous expression, 
  $\lambda$ is a dimensionless parameter, whose 
value could also be  large so to push the size of $\Lambda_{cc}$ towards  the gravitational cut-off of the theory.
 In the remaining of this section, for definiteness,  we will focus on the case in which $\lambda$ is positive. 
The homogeneous cosmological configuration is characterized by a 
de Sitter expansion: there are
  two available branches for the Hubble parameter (see eq. (\ref{H2brs}))
%
\bea
H^2_\mp&=&\frac{m^3}{18\,\beta_1\,M_{*} }\,\left[
{\lambda\mp\sqrt{\lambda^2-\frac{24 (1+3\gamma)}{(1+\gamma)^3}
}}
\right]\label{brhub}\,.
\eea
In discussing fluctuations, we have learned that scalar fluctuations do not propagate at quadratic level around this de Sitter space. Vector
fluctuations propagate as massless modes, while tensor fluctuations are described by a quadratic expansion of the Einstein Hilbert action,
with an induced renormalization of the effective Planck scale
 (see eqs (\ref{fmeff}) and (\ref{pmeff})):
\bea
M^2_{\mp}&=&\left( 1-\frac{24\,\left(1-\gamma\right)}{
\left(1+ \gamma\right)^3
\,\left( \lambda\mp\sqrt{\lambda^2-\frac{24(1+3\gamma)}{(1+\gamma)^3}
}
\right)^2
}\right)\,M_{*}^2\,.\label{renps2}
\eea
In the previous formula, 
a plus or minus sign depends on the choice for the branch of the Hubble parameter in eq (\ref{brhub}).  $M_{\mp}$ correspond
to the effective Planck mass  in the de Sitter solution of interest, while $M_\star$ is the original gravitational Planck scale around
a flat solution with no homogeneous vector profile. 

\subsection{Large bare cosmological constant}\label{sub-largeb}

These results have potentially interesting consequences for the scale of dark energy. 
Let us start by considering the case in which the numerical coefficient $\lambda$ in eq (\ref{defcc2})
  is positive and large
 (with `large' we mean $\lambda$ much bigger than $(1+3\gamma)/(1+\gamma)$)  
 corresponding to the case of potentially large vacuum energy $|\Lambda_{cc}|$.  

 We obtain for the two branches (\ref{brhub}) of the Hubble parameter the following limiting values
%
\bea
H^2_-&\simeq&\frac{2\,\left(1+3\gamma\right)\,m^3}{3\,\beta_1\,\left(1+\gamma\right)^3\,\lambda\,M_{*}}
\label{Hbm}\,,\\
H^2_+&\simeq&\frac{\lambda\,m^3}{9\,\beta_1\,M_{*}}\label{Hbp}
\,.\eea
In this large $\lambda$ limit, the effective Planck
mass for the two branches of solutions is 
\bea
M_{-}^2&\simeq&\frac{\left(1+ \gamma\right)^3\,\left(\gamma-1\right)\,\lambda^2}{6\,\left(1+3\gamma \right)^2}\,M_{*}^2 \label{imm}
+\frac{(3+\gamma)}{(1+3\gamma)}\,M_{*}^2\,,
\\
M_{+}^2&\simeq&M_{*}^2\,.\label{imp}
\eea

\smallskip

The simplest and most interesting case 
  to analyze corresponds to  the negative branch, $H_-$, $M_-$, and 
   $\gamma=1$.
   In this case, the parameter $\beta_2\,=\,0$, and eq (\ref{imm}) tells us  that the Planck mass does not get renormalized, $M_-\,=\,M_{\star}$. Eq (\ref{Hbm}) leads to  
  \bea
  H_-^2&=&\frac{m^3}{3\,\beta_1\,\lambda\,M_\star}
\\&=&\left(\frac{m^3}{3\,\beta_1}\right)^2
\,\frac{1}{\Lambda_{cc}}\label{slcc}
\\&=&\left(\frac{\Delta^3}{3\,\hat{\beta}_1}\right)^2
\,\frac{1}{\Lambda_{cc}}\label{tlcc}
  \eea
  In the second line of the previous expression, eq (\ref{slcc}), we used  eq (\ref{defcc2}):  interestingly, the Hubble scale results {\it inversely
  proportional} to the value of the bare cosmological constant. In the third line, we used the definition in (\ref{defratio}), showing that  
  the previous relation holds also in the limit of very small parameters $m$, $\beta_1$ in which we recover the Abelian and the (approximate) Galileon symmetries.  So, the larger is the bare cosmological constant $\Lambda_{cc}$, the smaller is $H_-^2$:
   the actual value of $H_-^2$  then depends on the strong coupling scale $\Delta$ controlling the Galileonic self-interactions of the 
   longitudinal vector polarization. 
   For 
  example, requiring that  $H_-$ is of order of present day  Hubble scale, 
  \be\label{todayH}
  H_{-}\sim 10^{-33} eV\,,
  \ee
  one finds 
  \be\label{vdsc}
  \Delta\simeq\left(3 \hat{\beta}_1\right)^\frac13\,\left( \frac{\Lambda_{cc}}{M_\star^4}\right)^{\frac16}   10^{7}\,eV\,.
  \ee
  If $\hat{\beta}_1\sim1$, and at the same time $\Lambda_{cc}$ is
  pushed towards the gravitational cut-off, the resulting $\Delta$
   is an intermediate scale between $H_-$ and $M_\star$, a 
   particularly interesting 
   situation since 
     in the limit 
   of small $m$, $\beta_i$ one recovers  
    approximate Abelian 
  and Galilean symmetries
   that protect the structure of our theory: see our discussion in Section \ref{sec-first}. 
   Different strong coupling scales $\Delta$, that might required by additional phenomenological considerations,
   can be obtained by changing the value of $\hat{\beta}_1$, 
     or reducing the size of $\Lambda_{cc}$ by considering a set-up  with supersymmetry broken well above the electroweak scale.  Also,
     notice that choosing a smaller $\Delta$ one can obtain a value for the Hubble parameter 
      induced by $\Lambda_{cc}$  much suppressed with respect to 
        its present day value, so to remove by this mechanism the contribution  of the bare cosmological constant, and
     then explain present day acceleration with some other option. 
  
  \smallskip
  
  The energy density stored in the vector field within this cosmological de Sitter solution is of the size of the bare cosmological constant:
    $\rho_A \sim \Lambda_{cc}$. Indeed,
    in this branch of solutions
      the constraint equation for $A_0$ automatically adjusts  the vector contribution to the energy momentum tensor,  in such a way 
     to compensate  to a large extent the bare cosmological  constant -- up to  the small leftover collected in the right hand side of (\ref{tlcc}).
   Analogously to the scalar-tensor theory of \cite{Charmousis:2011bf},
   we can avoid Weinberg's no-go theorem \cite{Weinberg:1988cp} because we are considering cosmological de Sitter configurations and a non-constant field  profile, while Weinberg's
     arguments apply only to situations in which  full Poincar\'e invariance  is maintained. Notice however that 
      the $vev$  of the  field $A_0$ itself can reach trans-Planckian values: using eqs (\ref{solfA0}) and (\ref{slcc}) one finds $A_0\sim \sqrt{\Lambda_{cc}}/m$. This is a feature shared with the scenario \cite{Batra:2008cc}.  It is not clear if it can constitute a problem though, since the field  $A_0$ is non-dynamical in our set-up~\footnote{We thank Tony Padilla for discussions on these points.}. 
  
  \bigskip
  
   Let us briefly recap our findings so far. We found that the structure of the tensor-vector action we started with admits
   a branch of de Sitter cosmological solutions in presence of a large bare cosmological constant $\Lambda_{cc}$, but with  a  
    value for the  Hubble parameter
     that is {\it inversely proportional} to a power of
     $\Lambda_{cc}$. 
      This result depends on the particular  set of operators we consider in our vector action,
       that 
       after solving for all the constraints 
       leads to a specific equation for the Hubble parameter 
       with the properties discussed above.
       The structure of our theory  can be argued to be technically natural, at least for the scales of interest and 
       in the limit   of small symmetry breaking parameters. 
%
        Hence, with the help of symmetry arguments  this set-up might  provide the opportunity 
         to 
          keep in a natural way the  
          dark energy scale  much smaller than the  cut-off of our  theory.

  \smallskip
  
  The very same arguments can be applied in the case in which $\beta_2$ is turned on, that is $\gamma\neq1$. However, in this case one has to take extra care to the renormalization of the Planck scale, see eq. (\ref{renps2})
  that can push  the effective gravitational cut-off, and the bare cosmological constant,  at   values much larger than $M_\star^4$, possibly destabilizing relations as (\ref{vdsc}): choosing a sufficient
  small (and technically natural) value for $\beta_2$, on the other hand, one can check that the previous arguments
  hold with little changes. 
  
  
\smallskip

While until now  we focussed on the interesting class  of solutions corresponding to the negative branch $H_{-}$
  let us briefly explain what happens for the positive branch
case. In the limit of large $\lambda$, the positive branch $H_+$ can still drive acceleration, but 
 in this case the value of the square of the Hubble parameter is {\it directly proportional} to the scale of the bare
 cosmological constant, see eq. (\ref{Hbp}). 
%
 %
  Hence this branch of solutions does not have much to say with respect to the cosmological constant problem,
and is  
    less interesting for the purpose of explaining present day acceleration.

  
 \subsection{Coupling with matter}
 
 While until this stage we focussed on the situation  of a universe filled with cosmological constant
 and vector fields, one can extend the  analysis of the previous subsection to the case in which other
 degrees of freedom are included. Since our vector theory breaks the Abelian symmetry, 
  non-standard 
   couplings  between the vector sector  and matter are allowed,
    that do not respect the gauge  symmetry: precisely those couplings can 
   be exploited for obtaining 
    a standard cosmological evolution in our set-up.
    This is a  subject that will be
  covered in full detail in \cite{toappear}. Nevertheless let us 
   start discussing this topic 
     here, by discussing   the easiest example
    of a  massless scalar field $\phi$ with no self-interactions, but that couples with the vector. For simplicity, in this subsection
   we focus our attention to the case $\beta_2\,=\,0$, that as discussed at the beginning of section \ref{sub-largeb}
    has the advantage of not  renormalizing  the Planck mass. 
   The kinetic term for the scalar Lagrangian is ${\cal L}_{m}(\phi,\,\partial\phi)=\frac12\partial_\mu \phi \partial^\mu \phi$.
  At the homogeneous level, the  energy density associated with such a  kinetic term is 
  $\rho\,=\,\dot{\phi}^2/2$.
  
   We allow for a
  direct, Planck suppressed coupling between the scalar and the 
  vector field, and include the following 
  contribution to the total action ${\cal S}$ of eq. (\ref{inact})
 \be\label{matact}
 {\cal S}_{m}\,=\,
 \int d^4 x\,\sqrt{-g}\,\left[ 
 -\frac{1}{2} \partial_\mu \phi \partial^\mu \phi+ \frac{\xi}{2\,M_\star^2}A_{\rho} A^{\rho} \partial_\mu \phi \partial^\mu \phi
 \right]
\,,
 \ee
 where
 $\xi$ is a dimensionless coupling, whose value
 will be specified later.
  The 
   new contribution (\ref{matact}) to the action 
   renders the background solution for the vector dependent on $\rho$. 
    Indeed,
   focussing again on configurations in which only 
   the $A_0$ component
     is turned on,
   the vector equations 
   of motion  are solved by 
  \be\label{A0pro}
  A_0\,=\,\frac{\left(m^2\,M_\star^2+\xi \rho\right)}{3\,\beta_1\,M_\star^2\,H}
\,.  \ee
 The  Friedmann equation governing the evolution of the scale factor receives contributions from the scalar
 energy density $\rho$ and becomes
 \be
 H^2\,=\,\frac{\Lambda_{cc}}{3\,M_\star^2}
 +\frac{\rho}{3\,M_{\star}^2}
 +\frac{\xi \,A_0^2\,\rho}{3 M_{\star}^4}
 +\frac{m^2\,A_0^2}{3 M_\star^2}-2\frac{\beta_1\,A_0^3\,H}{M_\star^2}
\,.
 \ee
 Plugging into the previous expression the profile (\ref{A0pro}) for $A_0$, and expanding for 
  large values of the bare cosmological constant $\Lambda_{cc}$ (as we
  did in the previous section \ref{sub-largeb}), we end with  the following expression for the Hubble parameter,
  that 
  generalizes
   the `negative branch'
  of solutions previously analyzed by including $\rho$:
 \be\label{negbs}
 H^2\,=\,\frac{m^6}{9\,\beta_1^2\,\Lambda_{cc}}
 +\frac{
 \left(  \xi\,m^4\,\Lambda_{cc}-m^6\, M_\star^2\right)
\,\rho}{9\,\beta_1^2\,M_\star^2\,\Lambda_{cc}^2}
+\dots
 \ee
 where the $\dots$ contain subleading corrections. The first  contribution to the right hand
 side is the  constant term 
 we found in eq. (\ref{tlcc}), inversely proportional to the bare cosmological constant.
  At this stage, it can  be more  conveniently expressed
 as
 \be
 \frac{m^6}{9\,\beta_1^2\,\Lambda_{cc}}\,\equiv\,\frac{\bar{\Lambda}}{3\,M_\star^2}
 \hskip0.5cm\to\hskip0.5cm
 m^6\,=\,\frac{3\,\beta_1\,\Lambda_{cc} \,\bar{\Lambda}}{M_\star^2}\label{defm6}
 \ee
 where the   quantity $\bar{\Lambda}$, in order
  not to curve excessively the space-time, 
  has to satisfy
 the inequality $\bar{\Lambda}\,\le\,\left(10^{-3}\, eV\right)^4\,\simeq\,10^{-120} \,M_\star^4$.  These 
  considerations   are equivalent to the ones     made around eqs.  (\ref{todayH}), (\ref{vdsc}).

 We can now choose  the  dimensionless parameter $\xi$ 
 appearing in the second term of (\ref{negbs}) 
  such  to obtain a standard form for
 the Friedmann equation. With this purpose we set
 \be
 \xi^3\,=\,\frac{3\,\beta_1^2\,
 \Lambda_{cc}
 \,M_{\star}^4}{9\,\bar{\Lambda}^2}
\, \left(1+\frac{\bar{\Lambda}}{\Lambda_{cc}} \right)^3\,. \label{defxi}\ee
Making  this choice, one finds the familiar expression for the Friedmann equation
\be
H^2\,=\,\frac{\bar{\Lambda}}{3\,M_\star^2}+\frac{\rho}{3\,M_\star^2}+\dots
\ee
plus contributions that are subleading in the limit $\bar{\Lambda}\ll\Lambda_{cc}$.
Hence we learn that by appropriately coupling matter with the vector field, as done in eq. (\ref{matact})
for this simple example of free scalar, one can reproduce standard cosmological evolution
at the homogeneous level. 

 The actual size
  of the dimensionless parameter 
  $\xi$ in eq. (\ref{defxi}) can be  chosen  smaller
   than unity so to break mildly the Abelian symmetry. In order to  achieve  this condition, we have to take very  
    small (but technically natural) values
 of $\beta_1$. Choosing a value for the bare $\Lambda_{cc}$ of the order of the gravitational
 cut-off scale, $\Lambda_{cc}\simeq M_\star^4$, and requiring (as explained
 above)  $\bar{\Lambda}\,\le\,10^{-120} \,M_{\star}^4$,
  the requirement of having $\xi\le1$ leads to the inequality $\beta_1\le 10^{-120}$. A very small number but -- as argued in the
  previous sections -- a  
  technically natural quantity in the 't Hooft sense,  since in the limit of small $m$, $\beta_1$ the theory recovers 
   Abelian and Galileon  symmetries.

  \subsection{Small bare cosmological constant}
  
  One can also be interested to what happens  in the limit of small $\lambda$, $\Lambda_{cc}$,
   to understand to what extent our Lagrangian  is able to drive cosmological acceleration exploiting 
  the dynamics of the vector degrees of freedom.
  It turns out that it is not possible to completely switch off 
  $\lambda$ and then $\Lambda_{cc}$. This since the stability of the  
  tensor sector requires a positive effective Planck mass squared
   $M_\pm^2$ for the two branches of solutions 
given  eq. (\ref{renps2}).  When demanding that  $M_{\star}^2\,>\,0$,
 this imposes the following condition for the two branches of solutions:
 \be
 \lambda\,\left( 
 \lambda\mp\sqrt{\lambda^2-\frac{24 (1+3\gamma)}{(1+\gamma)^3}}
 \right)\,\ge\,\frac{24}{\left(1+\gamma\right)^2}\,.
 \ee
 One can easily check that this inequality can  be satisfied only if $|\gamma|\ge1$.
 Imposing this condition 
 implies
 that   $(1+3\gamma)/(1+\gamma)^3>0$, and in order to have 
 a well definite square root in the previous formula we have to demand that  a non-vanishing $\lambda$ is turned on,
  such to satisfy
 \be
 \lambda^2\,\ge\,\frac{24 (1+3\gamma)}{(1+\gamma)^3}\,.
 \ee
 Imposing the minimal value of $\lambda$ that satisfies such inequality, we
 find the following values for the physical parameters
 \bea
 H_{\pm}^2&=&\sqrt{\frac{2\,(1+3\gamma)}{3\,\left(1+\gamma\right)^3}}\,\frac{m^3}{3\,\beta_1\,M_{\star}}
\,,
\\
 M_{\pm}^2&=&\frac{4\,\gamma}{1+3\,\gamma}\,M_\star^2
\,,
 \eea
 and when expressing the value of the Hubble parameter in terms of $\Lambda_{cc}$ we find
 \be
 H_{\pm}^2\,=\,\frac{2\,\gamma}{3\left(1+3\gamma\right)}
\,\frac{\Lambda_{cc}}{M_{\pm}^2}
\,.
 \ee
 Hence in this case the  vector field contributes  to the accelerated expansion changing 
by a factor $2\,\gamma/(1+3\gamma)$ the value of the effective cosmological constant. 
 
 \section{Conclusions}\label{sec-concl}

In this work, 
we studied    cosmological consequences of the vector-tensor theory   
  presented in \cite{Tasinato:2014eka}, 
 where the Abelian symmetry of a vector coupled to gravity is slightly broken by a mass
  term and ghost-free derivative self-interactions. 
  When studying cosmological expansion
   in the presence of large bare cosmological constant $\Lambda_{cc}$, we found that the theory admits  branches of
   de Sitter
    solutions in which
   the scale of the Hubble parameter is inversely proportional to a power of  $\Lambda_{cc}$. Hence,
     the larger  the bare cosmological constant is,
  the smaller the value of the Hubble constant, and parameters can be chosen in such a way that the resulting Hubble scale
  is of the order of the observed dark energy scale. 
   %
   %
     The theory   propagates up to three degrees of freedom
  in the vector sector: the two transverse vector components and a longitudinal scalar mode. In an appropriate
  limit, in which the symmetry breaking parameters are small, the theory recovers the Abelian symmetry plus an 
  additional Galileon symmetry acting on the longitudinal scalar mode.   
  The existence of the interesting branch of  cosmological solutions depends on the particular structure of our vector theory: exploiting the approximate 
  Galileon symmetries in the longitudinal sector, we argued that
 this structure can be stable for the energy  scales we are interested in. We also analyzed  the  dynamics of cosmological
 fluctuations around the  de Sitter  
    solutions, showing that 
     no manifest instabilities arise, and that the transverse vector polarizations become massless around  these  
     configurations.
        
   

\smallskip

We did not make explicit hypothesis on the nature of the vector field we considered:
 it could be the observed photon, or an additional light vector field not belonging to the Standard Model.  The main
 task left  to study is how
  normal matter gravitates in this cosmological set-up, and
  whether 
   it  
   couples with
   the vector field 
   in  a way  that does not spoil the features
   of the cosmological solutions we studied. 
     The longitudinal polarization of the 
     vector field 
     can mediate long-range forces, that can 
  nevertheless be screened by an analogue of the Vainshtein mechanism.  
   Normal matter, on the other hand, contributes to cosmological evolution  both  by directly curving the space-time, and by  modifying
   the homogeneous  time-dependent profile of the vector field when appropriate vector-matter couplings 
     are included.  
      A detailed study of these issues is left for future investigation.

 \acknowledgments
It is a pleasure to thank  Marco Crisostomi, Lavinia Heisenberg, Ivonne Zavala and especially  Tony Padilla for comments on the draft, as well as 
STFC for financial support through the grant ST/H005498/1.

\end{document}